\documentclass[aps,prb,twocolumn,groupedaddress]{revtex4}
\usepackage{graphicx}
\usepackage{amssymb}
\usepackage{amsmath}
\usepackage{xcolor}

\newcommand{\ket}[1]{|{#1}\rangle}
\newcommand{\bra}[1]{\langle{#1}|}
\newcommand{\braket}[2]{\langle{#1}|{#2}\rangle}
\newcommand{\abs}[1]{\left|{#1}\right|}
\newcommand{\MM}{\mathcal{M}}
\newcommand{\PPP}{\text{supp}}
\newcommand{\II}{\mathbb{I}}
\newcommand{\Hrel}{\mathcal{H}_{\Psi}}
\newcommand{\Irel}{\mathcal{\II}_{\Psi}}

\newcommand{\dd}{d}

\newcommand{\rhocond}{\rho_\text{cond}}

\newcommand{\SSS}{\mathcal{S}}

\newcommand{\D}{\mathcal{D}}

\newcommand{\jj}{\underline{j}}
\newcommand{\TPsi}{T_\Psi}

\newcommand{\Tr}{\mathrm{Tr}}

\newcommand{\trr}{{\tilde\rho}_\text{cond}}

\begin{document}

\title{Quantized recurrence time in iterated open quantum dynamics}
\author{P. Sinkovicz, Z. Kurucz, T. Kiss, J. K. Asb\'oth}
\affiliation{Institute for Solid State Physics and Optics, 
Wigner Research Centre, Hungarian Academy of Sciences, 
H-1525 Budapest P.O. Box 49, Hungary}
\date{\today}
\begin{abstract}

The expected return time to the original state is a key concept
characterizing systems obeying both classical or quantum dynamics. We
consider iterated open quantum dynamical systems in finite dimensional
Hilbert spaces, a broad class of systems that includes classical
Markov chains and unitary discrete time quantum walks on networks.
Starting from a pure state, the time evolution is induced by repeated
applications of a general quantum channel, in each timestep followed
by a measurement to detect whether the system has returned to the
original state.  We prove that if the superoperator is unital in the
relevant Hilbert space (the part of the Hilbert space explored by the
system), then the expectation value of the return time is an integer,
equal to the dimension of this relevant Hilbert space.  We illustrate
our results on partially coherent quantum walks on finite graphs.  Our
work connects the
previously known quantization of the expected return time for
bistochastic Markov chains and for unitary quantum walks, and shows
that these are special cases of a more general statement.  The
expected return time is thus a quantitative
measure of the size of the part of the Hilbert space available to the
system when the dynamics is started from a certain state. 

\end{abstract}
\maketitle

\section{Introduction}

An important task in physics is to observe the dynamics of systems and
predict their future behavior. Monitoring the evolution of a classical
system does not alter its dynamics in the ideal case; in quantum
mechanics, however, frequent measurements may have dramatic effects
due to the measurement back-action, e.g., freezing the dynamics, as in
the quantum Zeno effect \cite{quantum_Zeno,chandrashekar2010zeno}, or
losing coherence and thereby arriving at classical-like
behavior\cite{joos2003decoherence}. The problem becomes even more
complicated under realistic conditions, where the effects of the
environment cannot be neglected and the introduced noise affects the
fine quantum features needed for applications, e.g. in quantum
information\cite{lidar2013quantum}.

Discretization, both in time and space, is inherent in the definition
of many physical systems (e.g., networks), but can also occur as a
result of an approximation to make a system numerically tractable.
Iterations of a given generic time evolution step and assuming a
countable number of different states of the system thus represents a
large class of physical situations, including classical and quantum
networks\cite{perseguers2013distribution}.  A generic way to represent
a discretized iterative dynamical process, is a discrete time random
walk on a graph, in both the classical and the quantum case.

Discrete-time quantum walks (DTQW)\cite{venegas_review}, quantum
mechanical generalizations of random walks, have in the recent years
enjoyed increasing attention from both theoretical and experimental
\cite{PhysRevLett.104.050502,Karski09,alberti_electric_experiment,
  sciarrino_twoparticle,Broome10,PhysRevLett.104.100503,
  PhysRevLett.106.180403} physicists. The hallmark property of DTQWs
is that they spread faster than classical random walks: on a regular
graph, the variance of the position of the walker scales as
$\mathcal{O}(t^2)$ with the number $t$ of timesteps, rather than
$\mathcal{O}(t)$ as in the classical case.  This gives quantum search
algorithms using DTQWs\cite{Shenvi03} the same quadratic speedup as
possessed by the celebrated Grover algorithm\cite{Grover96} -- all the
more important since DTQWs can be used to implement universal quantum
computation \cite{Lovett2010}.
Characterization of quantum walks using fundamental concepts might not
only further our understanding of when and how the quantum speedup
arises, but can also lead to new types of algorithms based on DTQWs.

One of the important concepts used to characterize iterative dynamical
processes, such as random walks on graphs, is recurrence: whether a
system returns to its initial state, and if so, how long this return
takes. For finite, closed systems (where the dynamics conserves the
phase-space volume), the Poincar\'e theorem guarantees that recurrence
does take place, although the required time can be beyond the range of
any conceivable experiment.  The problem of this return
time in classical\cite{barreira2005poincare} and quantum systems
\cite{casati1999quantum} is an important question for many areas of
physics, from chaos theory to the microscopic foundations of
thermodynamics.
%

There is a broad class of classical iterative dynamical processes for
which the recurrence time, i.e., the expected return time to an
initial state $j$, turns out to be quantized, i.e., an integer
$T_j$. This is the class of bistochastic processes, for which the
completely mixed state is a stationary state\cite{Kemeny60}. In this
case, for each initial state $j$ the set $G_j$ of sites that are
reachable from $j$ by iterations of the timestep form an irreducible
component (all states in $G_j$, and only states in $G_j$ are reachable
from each other).  Since the process is bistochastic, in this
irreducible component the uniform distribution is a stationary state.
Consider now a trajectory of $N$ steps, started from state $j$, with
$N\to\infty$; the number of times site $j$ is visited tends to $N /
\abs{G_j}$, where $\abs{G_j}$ denotes the number of states in
$G_j$. The average return time to state $j$ is the average time delay
between such visits, which is $T_j= \abs{G_j}$.

Generalizing the concept of the first return time to iterative quantum
dynamical processes, Gr\"unbaum et al.\cite{Grunbaum13} have found a
striking fact: its expectation value is quantized. 
To define a first return time, they suggested that every timestep,
given by a unitary operator $U$, be followed by a measurement to
detect whether the walker has returned to the initial state or
not. Starting the system from a state $\ket{\Psi}$, they have found
that the expected return time $\TPsi$
is the dimension of the smallest eigenspace of $U$ that includes
$\ket{\Psi}$, which is always an integer number, or infinity. Thus,
the integer $\TPsi$ is a measure of the size of the system accessible
from the initial state, reminiscent of the classical case. This
similarity is all the more surprising given that the proof of
Gr\"unbaum et al.~is an intricate application of topological concepts
to quantum physics.

In the present work, we address the problem of the recurrence time for
all iterative \emph{open} quantum dynamical systems (IOQDS;
a.k.a.~quantum dynamical semigroups\cite{cirillo14}): a broad class of
processes that includes as special cases both the unitary case of
Ref.~\onlinecite{Grunbaum13} and classical Markov chains.  The
timestep of one period thus includes interaction with an environment,
which can be beneficial for transport\cite{Huelga08,Rebentrost09}. The
timestep is defined by an arbitrary quantum channel (trace preserving
completely positive map), represented by a superoperator $\SSS$, which
is followed by a measurement to decide whether the system has returned
or not. This iterated evolution can also be viewed as a generalized,
open discrete time quantum walk
\cite{attal2012open,lardizabal2014class}.  We prove that whenever the
timestep superoperator is \emph{unital}, i.e., whenever the completely
mixed state is a stationary state, $\SSS[\II]=\II$, the expected
return time $\TPsi$ to a pure state $\ket{\Psi}$ is an integer, equal
to the dimension of the part of the Hilbert space that the system
explores when started from $\ket{\Psi}$ (the relevant Hilbert space;
more precisely, we only need $\SSS$ to be unital in this relevant
Hilbert space).

This paper is structured as follows. In the next Section we fix the
notation for iterated open quantum dynamical systems, i.e.,
generalized DTQWs, also introducing the concepts of the conditional
density operators and of the relevant Hilbert space. In Section
\ref{sect:theorem} we prove the main statement of our paper, that the
expected return time for unital generalized DTQWs is equal to the
dimension of the relevant Hilbert space. In Section
\ref{sect:examples} we illustrate our results on DTQWs on finite
graphs. Finally, we provide a short outlook on consequences of our
results in Section \ref{sect:discussion}.

\section{First return time}
\label{sect:definitions}
We consider an iterated open quantum dynamical system, with dynamics
that can be fully or partially coherent.  The state is given by a
time-dependent density operator $\rho: \mathcal{H}\to \mathcal{H}$ in
a finite dimensional Hilbert space $\mathcal{H}$.  The dynamics takes
place in discrete time, $t\in\mathbb{N}$, starting from an initial
pure state $\ket{\Psi}$,
\begin{align}
\rho(0) &= \ket{\Psi}\bra{\Psi},
\end{align}
and generated by a fixed superoperator $\SSS$. To describe a real
physical process, $\SSS$ has to be trace preserving and completely
positive (TPCP), i.e., a quantum channel.  This is
equivalent by the Stinespring--Kraus representation theorem
\cite{kraus1983states} to the requirement that $\SSS$ can be written
in terms of a discrete set of $D$ Kraus operators $A_j: \mathcal{H}
\to \mathcal{H}$ as
\begin{align}
\rho(t+1) &= \SSS[\rho(t)] = \sum_{j=0}^{D-1} A_j \rho(t) A_j^\dagger.
\label{eq:kraus_intro}
\end{align}
The only restriction on the Kraus operators $A_j$ is the
normalization condition
\begin{align}
 \sum_{j=0}^{D-1} A_j^\dagger A_j  = \II, 
\end{align}
where $\II$ represents the unit operator on the whole Hilbert space
$\mathcal{H}$.

\subsection{Relevant Hilbert space} 

We can safely restrict our attention to the part of the Hilbert space
that the system explores, started from $\ket{\Psi}$ and undergoing the
iterations of $\SSS$.  This is the \emph{relevant Hilbert space}
$\Hrel$, which we define via the limit of the series of  projectors
\begin{align}
\label{eq:def_rel}
\Irel &= \lim_{n\to\infty} \PPP(\sum_{t=0}^n \rho(t)),
\end{align}
where $\PPP(\sigma)$ denotes the projector to the nonzero subspace
(support) of a Hermitian operator $\sigma$. The relevant Hilbert space
$\Hrel$ is the image of the operator $\Irel$ acting on
$\mathcal{H}$. 

The relevant Hilbert space $\Hrel$ is the smallest subspace of
$\mathcal{H}$ that contains the state $\ket{\Psi}$ and fulfils the
following property: for any positive semidefinite operator $\sigma:
\Hrel\to\Hrel$, we have $\PPP(\SSS[\sigma])\subseteq \Hrel$. In the
language of the Kraus operators of $\SSS$, as per
Eq.~\eqref{eq:kraus_intro}, this reads that for all $j$, and any
$\ket{\Phi}\in \Hrel$, we have $A_j \ket{\Phi} \in \Hrel$. These
statements, proved in Appendix \ref{app:relevant_hilbert}, ensure
that the restriction of the timestep operator $\SSS$ to the relevant
Hilbert space, defined as
\begin{align}
\SSS_\Psi[\rho] = \SSS[\Irel \rho \Irel] 
= \sum_{j=0}^{D-1} A_j \Irel \rho \Irel A_j^\dagger, 
\label{eq:S_Psi_def}
\end{align}
for $\rho:\mathcal{H}\to\mathcal{H}$, can be used in place of $\SSS$
as long as we only consider an iterative quantum dynamics started from
$\ket{\Psi}$. Since each of the Kraus operators map states from
$\Hrel$ into states in $\Hrel$, the Kraus operators of $\SSS_\Psi$ are
$A_j \Irel = \Irel A_j \Irel$.


The dimension of the relevant Hilbert space is equal to the trace
of the projector $\Irel$,
\begin{align}
\text{dim}(\Hrel) &= \text{Tr}(\Irel).
\end{align}
The relevant Hilbert space is guaranteed to be finite dimensional if
the full Hilbert space is finite dimensional.  

\subsection{Conditional dynamics} 

To define a first return time, we need to modify the dynamics and
monitor whether the system returns to the initial state or not.  At
the end of every timestep, we perform a dichotomic measurement, with
the projector corresponding to ``return'' given by
$\ket{\Psi}\bra{\Psi}$, and the projector of ``no return'' given by
its complement, which acts on a general density operator $\sigma$ as 
\begin{align}
\MM[\sigma] &= 
(\II-\ket{\Psi}\bra{\Psi}) 
\sigma
(\II-\ket{\Psi}\bra{\Psi}).
\label{eq:def_M}
\end{align}
The first return
time is the number of steps we need until we obtain ``return''. The
expected return time is the expectation value of this number.

A simple way to calculate the first return time in this monitored
system is to use a \emph{conditional density operator}
$\rhocond(t)$. This represents the state of the system under the
condition that it has never returned to the origin.  Using the
superoperator $\MM$ corresponding to ``no return'', 
this conditional density operator reads 
\begin{align}
\rhocond(t) &= 
(\MM \SSS)^t[\ket{\Psi}\bra{\Psi}] = 
(\MM \SSS_\Psi)^t[\ket{\Psi}\bra{\Psi}],
\label{eq:rhocond_def}
\end{align}
for $t\in \mathbb{N}$, including $\rhocond(0)=\ket{\Psi}\bra{\Psi}$.
The second equality above holds, since 
the projector $\MM$ does not lead outside the relevant Hilbert space:
for any positive semidefinite operator $\sigma$: $\Hrel\to\Hrel$, we
have $\PPP(\MM[\sigma]) \subseteq \Hrel$, with $\MM[\sigma]$ also
positive semidefinite (but possibly $\text{Tr } (\MM[\sigma]) < \Tr
\sigma$).

The trace of the conditional density operator $\rhocond(t)$ is the
probability $q_t$ that there was no return to the initial state up
until time $t$,
\begin{align}
q_t &= \Tr \rhocond(t).
\end{align}
We say that the dynamics is \emph{recurrent} if this quantity
converges to $0$.

\subsection{Expected return time} 

In this paper we are interested in the expected return time $\TPsi$,
i.e., the expectation value of the first return time,
\begin{align}
\TPsi &= \sum_{t=1}^{\infty} t p_t,
\end{align}
where $p_t$ denotes the probability that the first return to
$\ket{\Psi}$ happens at time $t$. We explicitly included the subscript
$\Psi$ in the expected first return time $\TPsi$, but we dropped it
from other quantities, e.g., the return probabilities $p_t$, and the
conditional probability density $\rhocond(t)$ for notational
simplicity. The return probabilities $p_t$ can be expressed in terms
of the probabilities $q_t$ of ``no return up until time $t$'' as
\begin{align}
p_t &= q_{t-1}-q_t, 
\label{eq:pt_def}
\end{align}
since the operator $\SSS$ preserves the trace. 
Using this, the
expected return time reads
\begin{align}
\TPsi = 1+ \sum_{t=1}^{\infty} q_t.
\label{eq:T_qq}
\end{align}

We can put Eq.~\eqref{eq:T_qq} into a very suggestive form using the 
conditional density operators, Eq.~\eqref{eq:trr_def}.
To do this, we define $\trr(t)$ as 
\begin{align}
\trr(t) &= \sum_{t'=0}^{t} \rhocond(t'). 
\end{align}
If the expected return time $\TPsi$ is finite, the series of operators
$\trr(t)$ converges, and we can define its limit as 
\begin{align}
\trr &= \lim_{t\to\infty} \trr(t) = \sum_{t=0}^{\infty} \rhocond(t) < \infty.
\label{eq:trr_def}
\end{align}
The expected return time $\TPsi$ reads simply 
\begin{align}
\TPsi = \Tr\,\trr.
\label{eq:T_q_rho}
\end{align}

We remark that the conditional density operators $\rhocond(t)$ span
the same subspace as the operators $\rho(t)$, 
\begin{align}
\label{eq:def_rel}
\Irel &= \PPP(\trr).
\end{align}
We relegate the proof of this statement to Appendix
\ref{app:relevant_hilbert}.

\section{First return time for unital iterated open quantum dynamical
 systems}
\label{sect:theorem}


We now come to the central result of this work, which can be written
succintly as
\begin{align}
\TPsi &= \Tr\, \Irel \quad \quad \text{if} \quad
\SSS[\Irel]=\Irel,
\label{eq:return_time_integer}
\end{align}
i.e., whenever the superoperator $\SSS$ defining a timestep is unital
on the relevant Hilbert space, the expected return time $\TPsi$ is an
integer, equal to the dimension of the relevant Hilbert space. In this
Section we prove this statement by showing that the operator $\trr$,
defined in Eq.~\eqref{eq:trr_def}, is a projector, i.e.,
\begin{align}
\trr &= \Irel. 
\label{eq:trr_theorem}
\end{align}
Eq.~\eqref{eq:return_time_integer} is a direct consequence of
Eq.~\eqref{eq:trr_theorem} and Eq.~\eqref{eq:T_q_rho}.

The proof of Eq.~\eqref{eq:trr_theorem} will be based on the
properties of the steady states of the conditional dynamics in the
relevant Hilbert space.  If a positive semi-definite operator $\sigma:
\Hrel \to \Hrel$ represents a steady state of the conditional
dynamics, it is necessarily $\sigma=0$. We will first show that
$\Irel-\trr \ge 0$, and then we prove that it is a steady state of the 
conditional dynamics.


\subsection{Unital iterated dynamics}

We say that an IOQDS, with timestep superoperator $\SSS$, started from
a pure state $\ket\Psi$ is $\Psi$-unital, if the restriction $\SSS_\Psi$ of
the operator $\SSS$ to the effective Hilbert space is unital, i.e., if
\begin{align} 
\SSS_\Psi[\Irel] &= \SSS[\Irel] = \Irel.
\label{eq:unitality_def}
\end{align} 
Thus a defining property of $\Psi$-unital IOQDSs is that the
completely mixed state in the relevant Hilbert space $\Hrel$ is a
steady state of $\SSS$.  In terms of the Kraus operators $A_j$ of
$\SSS$, $\Psi$-unitality is defined as the property
\begin{align}
 \sum_j A_j \Irel A_j^\dagger  = \Irel. 
\label{eq:unitality_def_kraus}
\end{align}
For an IOQDS, $\Psi$-unitality implies that the dual $\SSS_\Psi^\ast$
of the unital superoperator $\SSS_\Psi$, defined via its Kraus
decomposition as
\begin{align}
\SSS_\Psi^\ast[\rho] = \sum_{j=0}^{D-1} \Irel A_j^\dagger 
\rho 
A_j \Irel,
\label{eq:dual_kraus}
\end{align}
cf. Eq.~\eqref{eq:S_Psi_def}, is not only positive, but also preserves
the trace, and thus represents a valid physical operation.  We note
that since $\Irel$ is the projector to an invariant subspace, a
sufficient but not necessary requirement for an IOQDS to be $\Psi$-unital is
that
the superoperator $\SSS$ be unital, i.e., $\SSS[\II] = \II$. 

A useful property of unital TPCP superoperators $\SSS$ is that any
steady state of $\SSS$ is also a steady state of its dual,
\begin{align}
\SSS[\chi]=\chi \quad
&\Leftrightarrow \quad
\SSS^\ast[\chi]=\chi.
\label{eq:dual_steady}
\end{align}
This is a consequence of the nontrivial fact\cite{holbrook03} that for
any unital TPCP superoperator $\SSS$, all of its steady states $\chi$
commute with all of its Kraus operators $A_j$,
\begin{align}
\SSS[\chi]=\chi \quad 
&\Leftrightarrow \quad
\forall j: [A_j, \chi] = 0.
\label{eq:A_j_commute}
\end{align}

\subsection{First part of the proof: $\Irel-\trr$ is positive semidefinite}


We now prove that $\Irel-\trr$ is a positive semidefinite
operator. Since the support of $\trr$ is in the relevant Hilbert space
(as shown in Appendix \ref{app:relevant_hilbert}), this is equivalent
to the statement that all eigenvalues of $\trr$ are less than or equal
to 1. This on the other hand follows from the statement -- to be
proved below -- that for any time $t\in\mathbb{N}$ and any normalized
density operator $\sigma$ in the relevant Hilbert space $\Hrel$,
\begin{align}
\text{Tr } \left(\sigma \trr(t) \right) \le 1. 
\label{eq:trr_le_Irel1}
\end{align}
A corollary of this inequality is that for $\Psi$-unital IOQDSs with a finite
dimensional relevant Hilbert space, the series defining $\trr$
converges, and so, these dynamical processes are recurrent.

In order to prove Eq.~\eqref{eq:trr_le_Irel1}, we rewrite the overlap
of $\sigma$ and $\trr(t)$ as
\begin{align} \text{Tr} \left(\sigma \trr(t) \right) &= \sum_{n=0}^t 
\text{Tr} \left( \sigma
(\MM\SSS_\Psi)^n[\ket{\Psi}\bra{\Psi}]  \right),
\label{eq:sum_tr}
\end{align}
where we used $\SSS_\Psi$ instead of $\SSS$ by virtue of
Eq.~\label{eq:rhocond_def}.
Now for any two density operators $\sigma$ and $\rho$, we have 
\begin{align}
\text{Tr} \left( \sigma 
\MM[\SSS_\Psi[\rho]] \right) &=  
\text{Tr } \big( \sigma 
(\II-\ket{\Psi}\bra{\Psi}) \SSS_\Psi[\rho] (\II-\ket{\Psi}\bra{\Psi}) \big) \nonumber\\
\quad &=
\text{Tr} \left( \SSS_\Psi^\ast [\MM [\sigma]] 
\rho \right).
\end{align}
Applying this relation $n$ times to the $n$th term in the sum in
Eq.~\eqref{eq:sum_tr}, we obtain
\begin{align}
 \text{Tr } \left(\sigma \trr(t) \right) 
&=
\sum_{n=0}^t
 \bra{\Psi} (\SSS_\Psi^\ast \MM)^n[\sigma] \ket{\Psi}.
\label{eq:sum2}
\end{align}
This sum has a direct physical interpretation. Consider the filtered
dynamics defined by $\sigma(0)=\sigma$, and $\sigma(n+1) =
\SSS_\Psi^\ast \MM [\sigma(n)]$ for $n\in\mathbb{N}$.  Each term in
the rhs of Eq.~\eqref{eq:sum2} is the number by which the trace of the
conditional density operator $\sigma(n)$ decreases in each timestep
due to the projection applied at the beginning of the timestep. Thus,
this sum cannot exceed 1: at most, it is equal to $1$, in case the
state $\sigma(n)$ decays to $0$ under the iterations of
$\SSS_\Psi^\ast \MM$. This proves Eq.~\eqref{eq:trr_le_Irel1}, and, as
a consequence, recurrence of $\Psi$-unital IOQDSs, as well as the
relation $\trr \le \Irel $.

\subsection{Second part of the proof: $\Irel-\trr$ 
is a steady state of $\MM\SSS$, and thus, vanishes}

Let us now prove that the positive operator $\Irel-\trr$ is
proportional to a steady state of the
conditional timestep operator $\MM\SSS$,
\begin{align}
\MM\SSS[\Irel-\trr] &= \Irel-\trr.
\label{eq:eigL1trr}
\end{align}
We prove this equation by writing it as the difference of two
equations.  First, because of the unitality of $\SSS$ in the relevant
Hilbert space, we have
\begin{align}
\MM\SSS[\Irel] &= \MM[\Irel] = \Irel- \ket{\Psi}\bra{\Psi}. 
\label{eq:eigL1}
\end{align}
Second, because of the definition of $\trr$, Eq.~\eqref{eq:trr_def},
we have
\begin{align}
\MM\SSS[\trr] &= \trr - \ket{\Psi}\bra{\Psi}. 
\label{eq:eigLtrr}
\end{align}
Subtracting Eq.~\eqref{eq:eigLtrr} from 
Eq.~\eqref{eq:eigL1} gives 
Eq.~\eqref{eq:eigL1trr}. 

We now show that 
the conditional timestep operator $\MM\SSS$ can have no steady states
in the relevant Hilbert space: For all positive semidefinite Hermitian
operators $\chi: \Hrel \to \Hrel$,
\begin{align}
\MM\SSS[\chi] = \chi 
\quad &\Longrightarrow \quad
\chi = 0.
\label{eq:eigchi0}
\end{align}
To see this, consider a density operator $\chi$ that is a steady state
of $\MM\SSS$. For such a state, we must have $\Tr
(\MM\SSS[\chi])=\Tr\,\chi$, which is only possible if the projector
$(\II-\ket{\Psi}\bra{\Psi})$ does nothing to $\SSS[\chi]$.  Thus, $\chi$ is
not only a steady state of $\MM\SSS$, but of $\SSS$ as well,
\begin{align}
\chi &= \MM\SSS[\chi] = \SSS[\chi]. 
\end{align}
As a consequence, $\MM[\chi]=\chi$, and this can only hold if
\begin{align}
\bra{\Psi}\chi\ket{\Psi} &= 0.
\end{align} 
Now consider the overlap of $\chi$ with the density operators
$\rho(t)=\SSS^t[\ket{\Psi}\bra{\Psi}]$. For all $t\in \mathbb{N}$, we obtain
\begin{align}
\text{Tr }\left(\chi \SSS^t[\ket{\Psi}\bra{\Psi}] \right) &= 
\bra{\Psi} (\SSS^\ast)^t [\chi] \ket{\Psi}=\bra{\Psi}\chi\ket{\Psi}=0
\end{align}
where we used Eq.~\eqref{eq:dual_steady}.  Since the eigenvectors of
$\rho(t)$ span the relevant Hilbert space, this proves that $\chi=0$. 

Combining Eq.~\eqref{eq:eigL1trr} with the statement
\eqref{eq:eigchi0}, we have $\Irel-\trr = 0$, which amounts to the
theorem we set out to prove, Eq.~\eqref{eq:trr_theorem}.  

\section{Examples}
\label{sect:examples}

Having proved that for $\Psi$-unital IOQDSs, the expected return time
is equal to the dimension of the relevant Hilbert space,
Eq.~\eqref{eq:return_time_integer}, we next illustrate the statement
on a few examples. In all of these, the timestep superoperator $\SSS$
is obtained by concatenating two quantum channels: a fully coherent
channel, defined via a unitary timestep operator $U$, followed by an
incoherent quantum channel, whose Kraus operators we will denote by
$B_j$. This construction allows us to controllably break unitarity,
symmetry, and $\Psi$-unitality of the IOQDS.

\subsection{Uniform decoherence on star graphs}

\begin{figure}[!htb]
  \includegraphics[width=0.6\columnwidth]{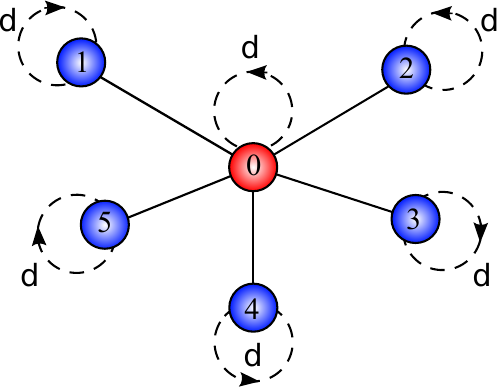}
  \caption{(Color online) A star graph of $M=6$ nodes, on which a quantum
    walk is started from the central node, 0. Each timestep consists
    of a coherent part, via a Hamiltonian, Eq.~\eqref{eq:def_U_H},
    with hopping along the continuous lines, followed by uniform
    decoherence with rate $d$, per Eq.~\eqref{eq:uniform_decoherence}.
    Although the full Hilbert space is 6 dimensional, in the absence
    of decoherence, the relevant Hilbert space is spanned by just 2
    states, as explained in the text. }
\label{fig:graph_1}
\end{figure}

Our first example is a quantum walk on a star graph of $M$ nodes (or
sites), as shown in Fig.~\ref{fig:graph_1}. 
The unitary part of the timestep defined via a
Hamiltonian as
\begin{align}
U &= e^{-i H};& \quad 
H &= \sum_{j=1}^{M-1} v_{j} \ket{j}\bra{0} + h.c.,
\label{eq:def_U_H}
\end{align}
where $v_j\neq 0$ are arbitrary nonzero complex hopping amplitudes.
During each timestep, a unitary operation by $U$ is followed by a
decoherence process $\D$ of rate $\dd$, i.e., a suppression of the
off-diagonal elements of the density matrix in the preferred basis
given by the nodes of the graphs,
\begin{equation}
\left( \D[\rho] \right)_{m,n} 
=\begin{cases}
    \rho_{m,n}, & \text{if $m=n$},\\
    (1-\dd) \rho_{m,n}, & \text{otherwise}.
  \end{cases}
\label{eq:uniform_decoherence}
\end{equation}
The superoperator
$\SSS$ for one complete timestep reads
\begin{equation}
\SSS[\rho] = \D [U \rho U^\dagger], 
\end{equation}
Tuning $\dd$ allows us to control the degree of decoherence, from
fully coherent time evolution ($\dd=0$), to full decoherence
($\dd=1$). In the latter case the dynamics can be given as a classical
Markov process.  

The decoherence channel $\D$ also has a
representation in terms of $M$ Kraus operators $B_j$, which read
\begin{subequations}
\begin{align}
B_j  &= \sqrt{\dd} \ket{j}\bra{j}, \quad \text{for } j=0,\ldots,M-1;\\
B_M  &= \sqrt{1-\dd}\, \II.
\end{align}
\end{subequations}

To gain a more intuitive understanding of the dynamics, we can rewrite 
the Hamiltonian as 
\begin{align}
H &= \bar{v} \ket{v} \bra{0} + h.c., 
\label{eq:star_ham}
\end{align}
with 
\begin{align}
\bar{v} &= \left(\sum_{j=1}^{M-1} \abs{v_j}^2 \right)^{1/2};&\quad
\ket{v} &= \bar{v}^{-1} \sum_{j=1}^{M-1} v_j \ket{j}.
\end{align}
Thus, 
it has only 2 eigenstates with
support on $\ket{0}$, namely,
\begin{align}
\ket{\pm} &= \frac{1}{\sqrt{2}} \left( \ket{0} \pm
\ket{v}\right). 
\label{eq:eig_state_pm}
\end{align}
That all other eigenstates have no overlap with $\ket{0}$ is clear
because $\abs{\braket{0}{+}}^2+\abs{\braket{0}{-}}^2 = 1$. The other
eigenstates form a subspace of 0 energy, spanned by the
unnormalized and nonorthogonal, but linearly independent set of
vectors $\ket{\Psi_j}$, with $j=1,\ldots,M-2$, defined as
\begin{align}
\ket{\Psi_j} &=   
\sum_{l=1}^{M-1} \frac{e^{i jl 2\pi/M}}{v_l^\ast} \ket{l}.
\label{eq:dark_states}
\end{align}
The states $\ket{\Psi_j}$ are dark states: from these states,
destructive interference between the hopping processes in the
Hamiltonian prevent the system from getting to $\ket{0}$ during the
unitary part of the timestep.

In the fully coherent case, defined as $\dd=0$, the eigenstates of the
Hamiltonian $H$ are also steady states of the quantum walk. Since only
two of these eigenstates have overlap with the initial state of the
walk, $\ket{0}$, the relevant Hilbert space, spanned by $\ket{+}$ and
$\ket{-}$, has dimension 2. Thus, in this fully coherent, unitary
quantum walk, the expected return time is $T_0 = \sum_{t=0}^\infty t
p(t) = 2$.

If the decoherence rate $\dd$ is nonzero, the states $\ket{\Psi_j}$ of
Eq.~\eqref{eq:dark_states} are no longer dark states, as the
destructive interference isolating them from $\ket{0}$ is no longer
complete.  Thus, the relevant Hilbert space becomes the whole Hilbert
space (no transition amplitude is 0), and the expected return
time is equal to the number of nodes on the graph, $T_0=M$.

The fact that the expected return time is independent of the unitary
operator $U$, as long as all hopping amplitudes $v_j$ are nonzero, can
be surprising, given that the probability distribution of the return
times $p_t$ is quite sensitive to the choice of $U$. We illustrate
this in Fig.~\ref{fig:exampleB} on two random examples (for details on
the numerical method, see Appendix \ref{app:numerics}) with a graph
consisting of $M=6$ nodes.  It is certainly not evident to the naked
eye, but confirmed by the simulations, that the expectation value of
the return time for the examples shown is $T_0=2$ without decoherence,
and $T_0=6$ with decoherence.

\begin{figure}[!htb]
  \includegraphics[width=\columnwidth]{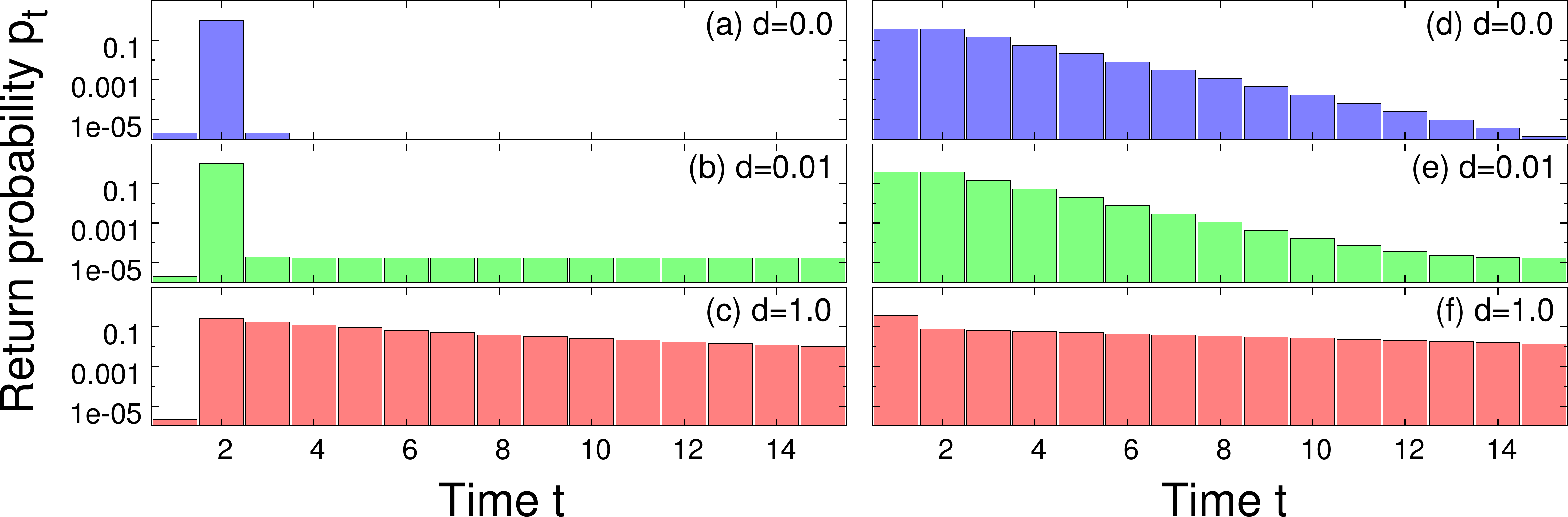}
  \caption{(Color online) Probability distributions of the first
    return time on a star graph of degree 6, for two different sets of
    random hopping amplitudes (left column, right column), and
    increasing decoherence rates (from top to bottom). The expected
    return time with no decoherence (top row) is $T_0=2$, with
    decoherence (middle and bottom row) it is $T_0=6$.}
\label{fig:exampleB}
\end{figure}

\begin{figure}[!htb]
  \includegraphics[angle=0,width=\columnwidth]{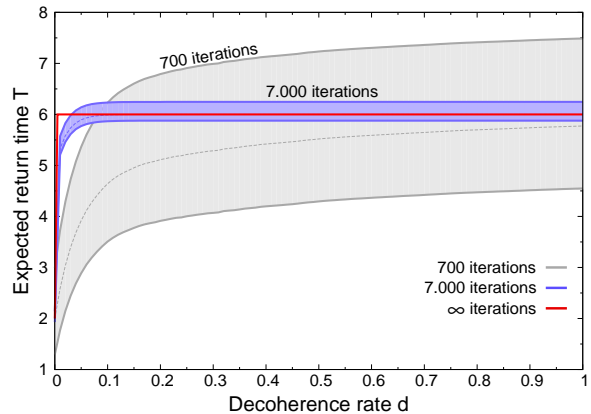}
  \caption{(Color online) First return times after $L$ timesteps on a
    star graph of degree 6, as functions of decoherence rates $d$.
    The shaded areas are between the lower and upper decile, for
    $L=700$, $L=7000$, and $L=\infty$; the medians are shown with
    dashed lines.}
\label{fig:convergence}
\end{figure}

To understand how even an infinitesimal amount of decoherence can
change the expected return time to $T_0=6$ from $T_0=2$, we explore the
partial expected return time $T_0^{(L)}$, i.e., the expected return time
after a finite number $L$ of timesteps.  This quantity is defined by
\begin{align}
T_0^{(L)} &= \sum_{t=1}^L t p_t + (L+1) (1-\sum_{t=1}^L p_t), 
\label{eq:TL_def}
\end{align}
where the DTQW is done for $L\in\mathbb{N}$ timesteps only, and if the
walker does not return, it is assumed to return in timestep $L+1$.
Although the expected return time $T_0$ does not depend on the hopping
amplitudes $v_j$, the quantities $T_0^{(L)}$ do. To show the extent of
this dependence, we sample the hopping amplitudes $v_j$ uniformly on
the complex disk of unit radius and plot the median, the upper decile,
and the lower decile of the distribution of the return times after $L$
timesteps $T_0^{(L)}$ in
%
Fig.~\ref{fig:convergence}. 
As the number $L$ of observed
timesteps increases, the range of the expected return
times goes down, and the expected return times approach the
asymptotic value.


\subsection{Breaking unitality: population transfer 
processes on complete graphs}
\label{subsect:breaking_unitality}

\begin{figure}[!htb]
  \includegraphics[width=0.6\columnwidth]{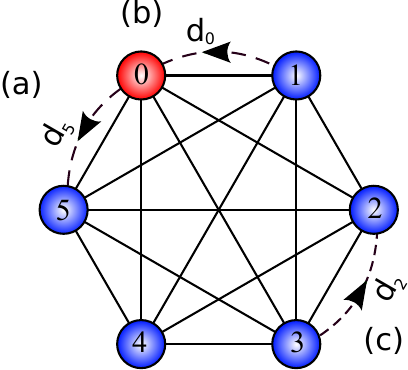}
  \caption{(Color online) A fully connected graph of $M=6$ nodes, on
    which a quantum walk is started from node 0. Each timestep
    consists of a coherent part, defined via a fixed random unitary
    operator (chosen uniformly from a Circular Unitary Ensemble),
    followed by an incoherent process that transfers population
    between two fixed sites, along a dashed line, as in
    Eq.~\eqref{eq:kraus_poptransfer}. We consider three examples:
    where this population transfer delays the return (a), where it
    speeds up the return (b), and where it is neutral (c).}
\label{fig:graph_2}
\end{figure}

In a next set of examples we break unitality of a DTQW on a fully
connected graph of $M$ nodes, in a controlled way. 
This is achieved using an asymmetric partial population transfer
process that follows the coherent part of the timestep. If this
population transfer is from a fixed ``source'' site to a fixed
``target'' site, it creates an accumulation of probability at the
target site, and thus, is not unital. As a consequence, the expected
return time will not be an integer, and will depend on the system
parameters in a continuous way. 

The incoherent partial population transfer from one fixed
source site ($j+1$) to another fixed target site ($j$), with rate
$\dd_j$, is defined via its Kraus operators $B_0^{(j)}$ and
$B_1^{(j)}$, as
\begin{subequations}
\begin{align}
B_{0}^{(j)}  &= \sqrt{\dd_j}\, \big{|} \, {j} \,\big{\rangle} \big{\langle} 
\,(j+1) \,\text{mod}\,M \, \big{|};\\
B_{1}^{(j)}  &= \II + (\sqrt{1-\dd_j}-1) \ket{j}\bra{j}.
\end{align}
\label{eq:kraus_poptransfer}
\end{subequations}
The full timestep operator $\SSS$ of the DTQW of this Section consists
of a unitary part, followed by partial population transfer,
\begin{align}
\rho(t+1) &= \SSS^{(j)}[\rho(t)]
= \sum_{l=0,1} B_{l}^{(j)} U \rho(t) U^\dagger B_l^{(j) \dagger} 
\end{align}

To study the effect of the asymmetric population transfer numerically,
we used a fully connected graph of $M = 6$ nodes, as shown in
Fig.~\ref{fig:graph_2}. There are 3 inequivalent ways of choosing the
source and target nodes for the extra incoherent population transfer
process, indicated by (a,b,c) in Fig.~\ref{fig:graph_2}.  In all of
these cases, the population transfer breaks unitality, the expected
return time $T_0$ can deviate from the number of nodes, $M$, and
depends on the unitary operator $U$.  To characterize this dependence,
in each case we numerically determined the distribution of the
expected return time $T_0$. We generated 2000 random instances of the
operator $U$, picked uniformly from the set of unitary operators on
the $M$-dimensional Hilbert space, using the circular unitary
ensemble\cite{CUE07}. For each value of the population transfer rate
$d$, we calculated the median, and the upper and lower deciles of the
distribution of the expected return time $T_0$.

\begin{figure}[!h]
 \includegraphics[angle=270,width=0.49\columnwidth]{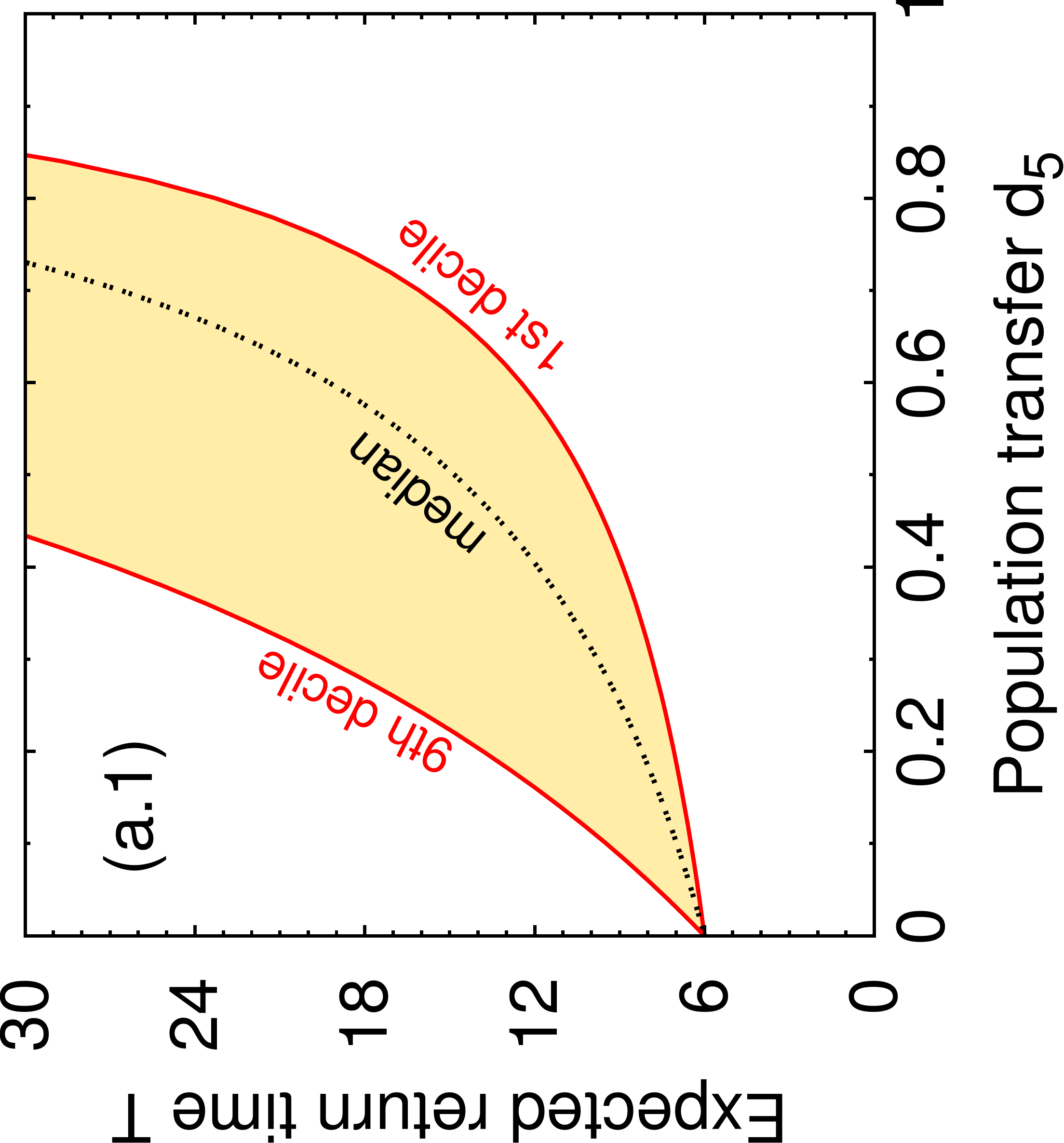}%
 \includegraphics[angle=270,width=0.49\columnwidth]{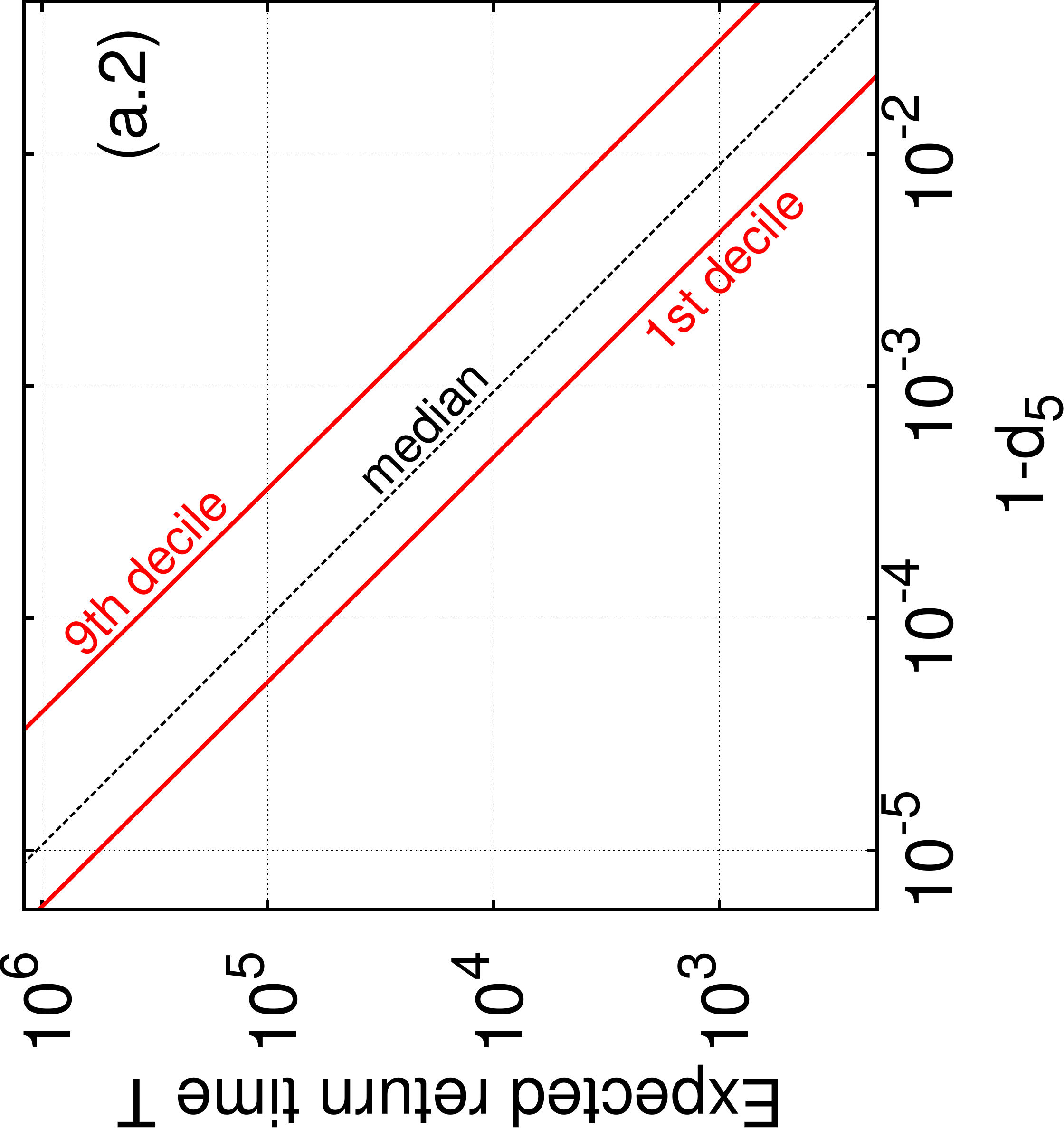}
 \includegraphics[angle=270,width=0.49\columnwidth]{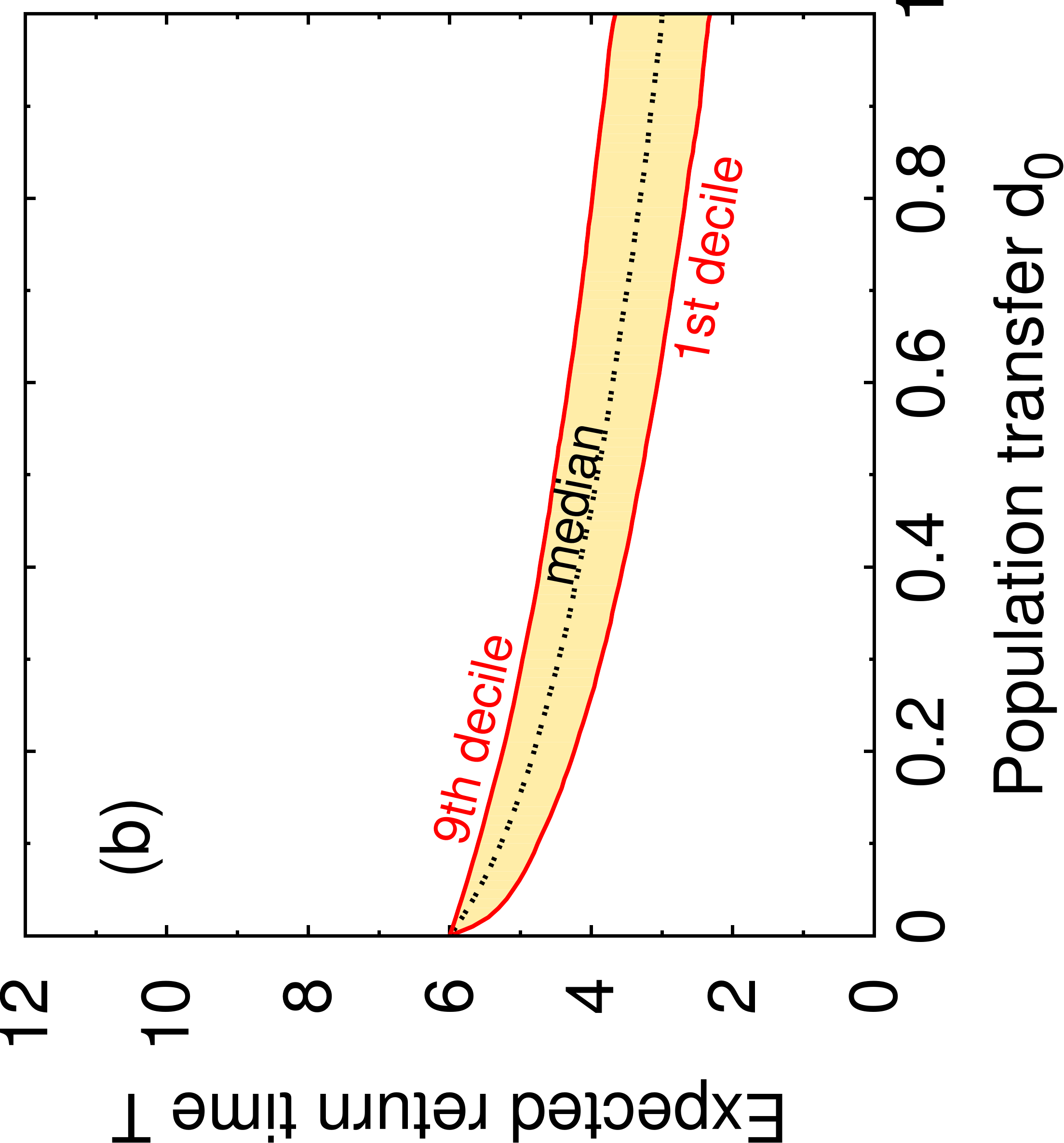}%
 \includegraphics[angle=270,width=0.49\columnwidth]{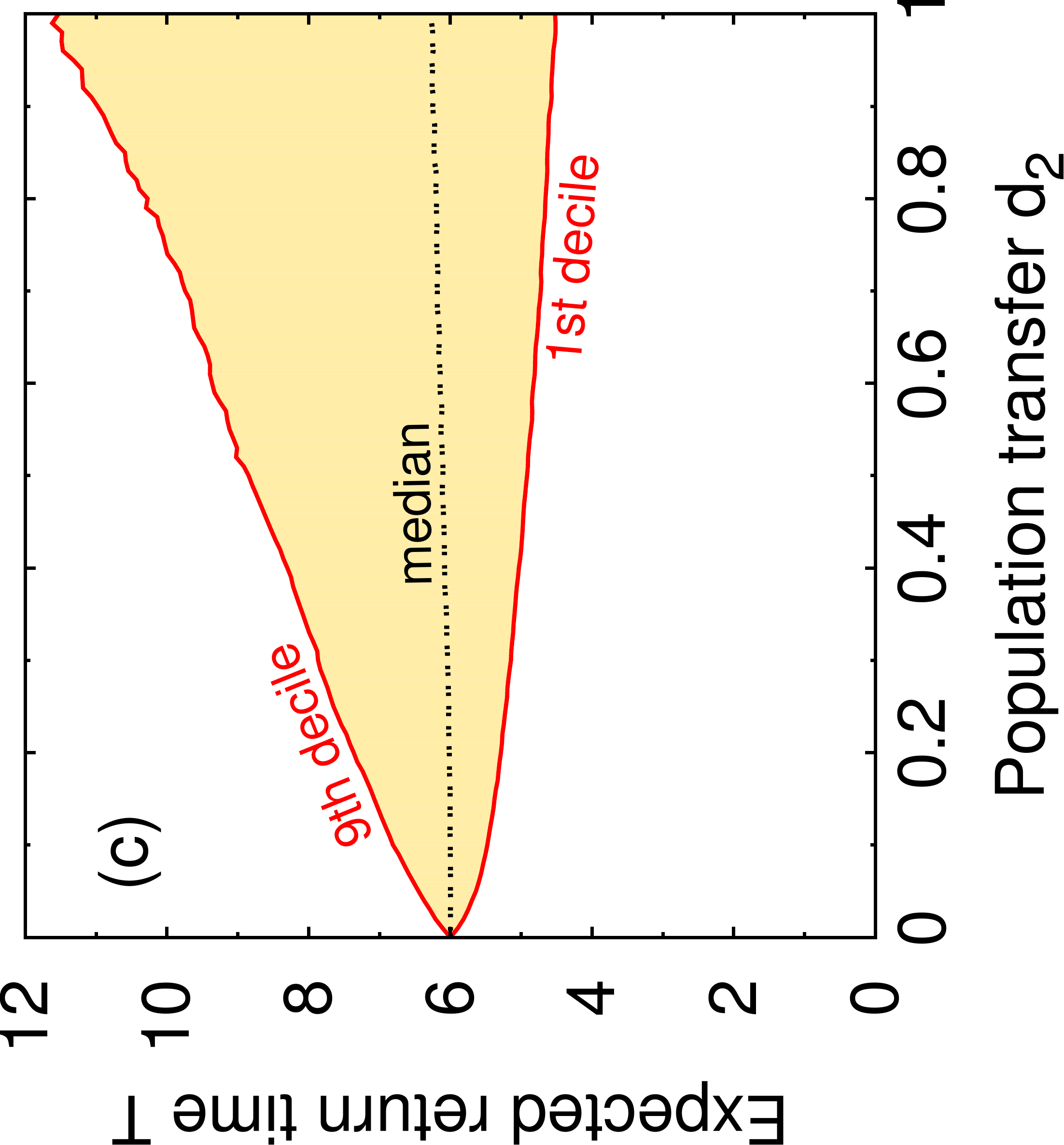}
 \caption{(Color online) The expected return time $T_0$ for a DTQW on a
   graph of 6 nodes, where a unitary operation is followed by
   asymmetric partial population tranfer towards the initial
   state. The unitary operator is picked at random from the circular
   unitary ensemble; thus, $T_0$ lies in the typical range shown by the
   shaded area, between the upper and lower decile (continuous lines),
   with the median also shown (dashed lines).  If the population
   transfer is away from the origin, (a.1) and (a.2), the expected
   return times increase as a function of the population transfer
   rate. If it drives the walker back to the origin, (b), the expected
   return times decrease.  For extra population transfer between two
   of the unobserved sites (c), the expected return time depends on
   the transfer rate. }
 \label{fig:asymmetric_transfer}
\end{figure}

Our numerical results, shown in Fig.~\ref{fig:asymmetric_transfer},
confirm that the asymmetric population transfer induces a spread of
the expected return times. Moreover, depending on its direction, the
population transfer can also change the average (median) of the return
time.  When the transfer is directed away from the initial state
(i.e., the target site is $j=5$, case (a) in Fig.~\ref{fig:graph_2}),
the expected return times increase, as shown in
Fig.~\ref{fig:asymmetric_transfer}(a).  In the limit $\dd\to 1$, the
expected return time diverges as $T_0 \propto 1/(1-d)$, as it would in a
classical walk. When the incoherent process drives the walker back
towards the initial site ($j=0$, case (b) in Fig.~\ref{fig:graph_2}),
the average of the
expected return time decreases as a function of the transfer rate, 
as shown in Fig.~\ref{fig:asymmetric_transfer}(b). In
the limit $\dd \approx 1$, we find that the median is approximately
half the number of sites, $M/2$, which matches the intuition that in
this case, 2 out of $M$ sites correspond to successful
return. Finally, for neutral population transfer ($j=2$, case (c) in
Fig.~\ref{fig:graph_2}), , as shown in
Fig.~\ref{fig:asymmetric_transfer}(c), the expected return time $T_0$
acquires a spread due to the population transfer, but the median stays
approximately independent of the population transfer rate.



\subsection{Breaking detailed balance but not unitality: Population transfer in a loop}

\begin{figure}[!htb]
  \includegraphics[width=0.5\columnwidth]{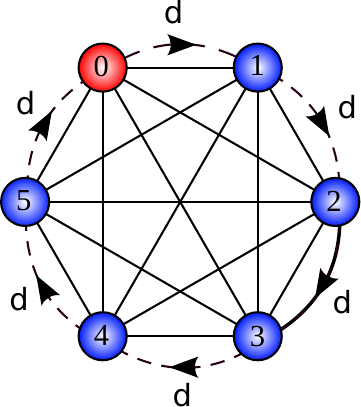}
  \caption{(Color online) A fully connected graph of 6 nodes, on which
    a quantum walk is started from node 0. Each timestep consists of a
    coherent part, defined via a Hamiltonian with random hopping
    amplitudes, followed by an incoherent process that transfers
    population in a loop along the dashed lines, as in
    Eqs.~\eqref{eq:ss_loop_poptransfer} and
    \eqref{eq:kraus_loop_poptransfer}.}
\label{fig:graph_3}
\end{figure}

Our final numerical example shows that asymmetric population transfer
does not necessarily break unitality. We consider the population
transfer to take place along a closed directed loop with uniform
transfer rate $d$, as shown in Fig.~\ref{fig:graph_3}, such that it
induces a probability current. In that case, there is no detailed
balance in the system, but the population transfer channel is unital,
and so the expected return time does not deviate from the quantized
value.

Each timestep consists of a unitary operation followed by the
incoherent transfer, 
\begin{equation}
\SSS[\rho] = \sum_{j=0}^M B_j U \rho
U^\dagger B_j^\dagger.
\label{eq:ss_loop_poptransfer}
\end{equation}
The Kraus operators $B_j$ are defined as 
\begin{subequations}
\begin{align}
B_j  &= \sqrt{\dd} \ket{j}\bra{j+1} \quad \text{for } j=0,\ldots,M-2;\\
B_{M-1}  &= \sqrt{\dd} \ket{M-1}\bra{0};\\
B_M  &= \sqrt{1-\dd} \,\II.
\end{align}
\label{eq:kraus_loop_poptransfer}
\end{subequations}
Since $\sum_{j=0}^M B_j B_j^\dagger =\II$, the whole timestep
operator $\SSS$ is unital, and so the expected return time will be
$M$, as confirmed by our numerics.

\begin{figure}[!h]
 \includegraphics[angle=0,width=0.95\columnwidth]{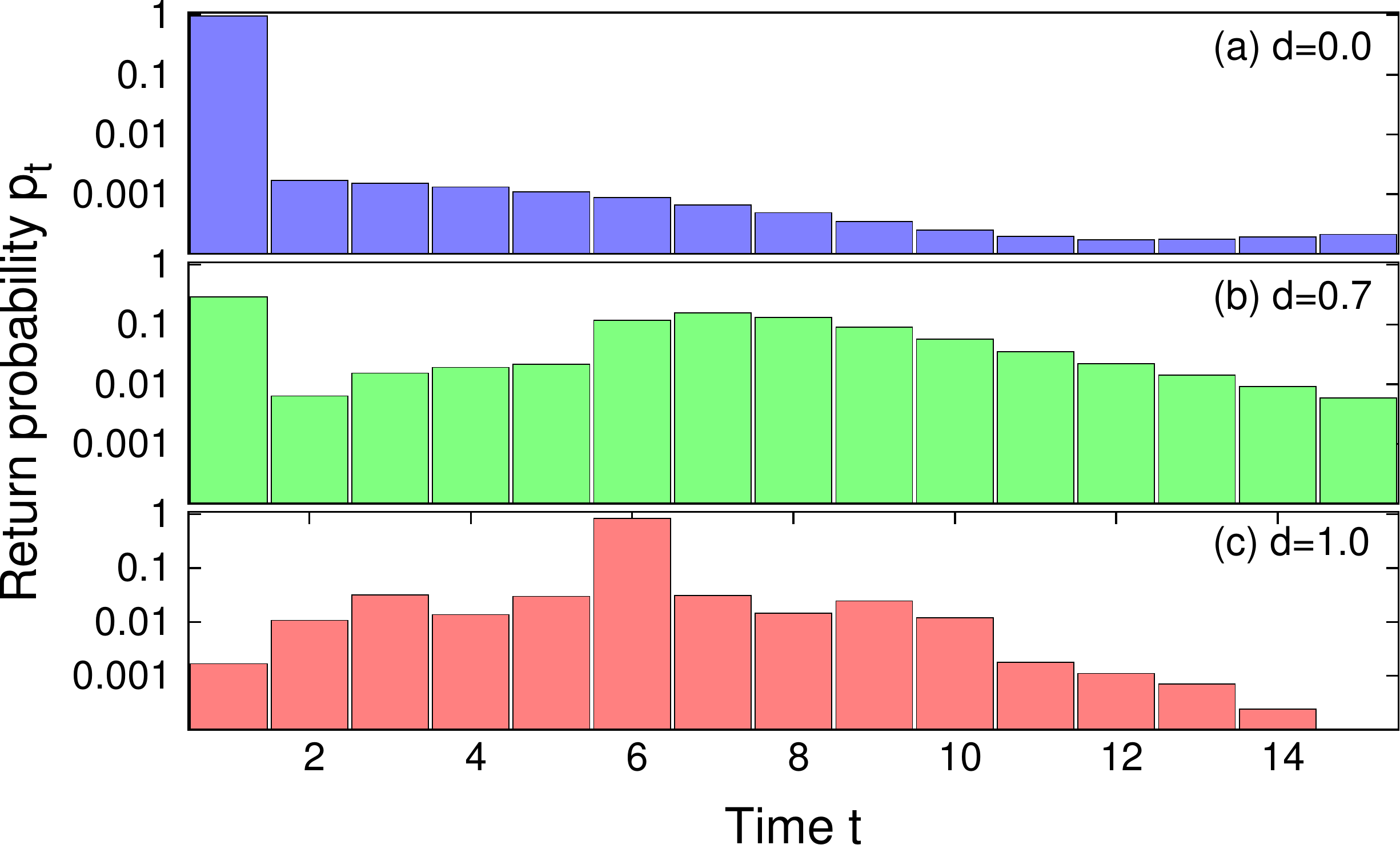}
  \caption{(Color online) Probability distributions of the first
    return time of a DTQW on a graph of degree 6, where a unitary
    operation is followed by an incoherent population transfer in a
    loop that includes all the sites. For the unitary part of the
    walk, we use $U=e^{-iH}$, with $H$ is a random Hermitian matrix
    with all matrix elements $\abs{H_{lm}}<0.1$. Although the
    distributions are qualitatively different, their expectation
    values are all equal, $T_0=\sum_t t p_t = 6$, because the walks are
    unital.  }
 \label{fig:loop}
\end{figure}

If the unitary operator $U$ is close to unity, it is worthwhile to
look at the probability distribution of the return time $p_t$, since
there is an interesting effect.  $U=e^{-iH}$, where $H$ is a
Hamiltonian with random hopping amplitudes uniformly distributed on
the disk of radius $0.1$. In this case, almost no transitions happen
during the coherent part of the timestep. Thus, for $\dd\approx 0$, we
have $p_1\approx 1$, and the expected return time is $T_0=M$ only
because of the exponential tail of the distribution. If the rate is
$\dd\approx 1$, however, the walker is most likely taken on a round
trip by the population transfer process, and so we obtain a peak in
the distribution at $p_M\approx 1$.  
Fig.~\ref{fig:loop}, 

\section{Discussion}
\label{sect:discussion}

We proved that in iterated open quantum dynamical systems, unitality
of the time evolution superoperator warrants that the expected 
return time is quantized.  We introduced the concept of the relevant
Hilbert space, which is spanned by the states of the system that can
be reached from a given initial state, and proved that its dimension
gives the expectation value of the first return time.  Our work treats
a broad class of physical systems on the same footing, including -- as
limiting cases -- fully coherent iterated quantum dynamical systems
(quantum walks), as well as classical Markov chains.

An immediate question raised by our work is: what about the expected
return time in iterated quantum dynamical systems where the timestep
superoperator is not unital?  In the fully coherent case, this
question does not arise, as the timestep superoperator can be
constructed from a single unitary Kraus operator, and is thus always
unital.  In the fully classical limit, there is a well known answer to
this question, given by the Kac lemma\cite{Kac47}: the expected return
time $T_j$ is the inverse of the maximum of the weight of the initial
state $j$ in an equilibrium distribution (the maximum taken over all
possible equilibria).
Detailed analysis of our numerics, e.g., the data processed for
Section \ref{subsect:breaking_unitality}, suggests that a statement
analogous to the Kac lemma might hold for iterated open quantum
dynamical systems. For an analytical treatment, however, more
theoretical tools are needed, just as for the treatment of the
recurrence to a more general initial condition, e.g., a subspace
spanned by a set of initial states\cite{bourgain2014quantum}.

As is often the case with classical concepts, the generalization of
the notions of recurrence, and of the expected return time, from
random walks to quantum walks is not unique. Besides the approach we
take in this paper, there is an alternative route, an ``ensemble
approach'', useful to obtain estimations for efficiency of quantum
protocols. This consists in letting the quantum walk run undisturbed
and after a fixed time measure the position distribution of the
walker\cite{SJK08}, or its full quantum
state\cite{chandrashekar2010fractional}. The expected return time,
defined in this way, does not necessarily take on an integer value
even in the fully coherent case: it can exceed or stay below the
dimension of the Hilbert space\cite{Grunbaum13}.

Recurrence is not only interesting for iterated open quantum dynamical
systems, but for continous time processes as well, whose time
evolution is prescribed by a quantum master equation. Here, to define
a first return time, the time evolution is considered punctuated by
measurements to detect the return of the walker. If these measurements
are randomly timed, according to a Poisson process, the hitting times
can become infinite even in the unitary case\cite{Varbanov08}; no
simple picture for the value of the return time has been yet found.
The measurements can also be regularly timed: in this case, we obtain
a continuous-time realization of the iterated open quantum dynamics,
and our results considering the return time apply. In this latter
case, it should be possible to cast the requirement of unitality, as
well as the dimension of the relevant Hilbert space, in a simple
formula for the Lindblad operators of the master equation.  As an
aside, there is a continuous-time generalization of the ensemble
approach, with measurements that are either randomly distributed or
regularly timed\cite{darazs10}.

Our results give a concrete quantitative measure of 
the size of the part of the Hilbert space accessible from
$\ket{\Psi}$. This could be a useful tool in the analysis of complex
quantum networks\cite{faccin13,faccin14}. The expected return time
$\TPsi$, or, for a more detailed picture, the partial expected return
times defined in Eq.~\eqref{eq:TL_def}, can be locally measured even
with limited access to the full quantum network.



\begin{acknowledgements}

We thank A.~G\'abris and J.~Novotny for useful discussions.  We
acknowledge support by the Hungarian Scientific Research Fund (OTKA)
under Contract Nos. K83858, NN109651 and the Hungarian Academy of
Sciences (Lend\"ulet Program, LP2011-016).

\end{acknowledgements}

\appendix

\section{Relevant Hilbert space}
\label{app:relevant_hilbert}

In this Section we prove some of the properties of the relevant
Hilbert space used in the paper.

Take a Hilbert space $\mathcal{H}$, and take any superoperator $\SSS$,
defined by its effect on density operators $\rho: \mathcal{H} \to
\mathcal{H}$, via the Kraus operators $A_j$ as
\begin{align}
\SSS[\rho] = \sum_{j=1}^{D} A_j \rho A_j^\dagger.  
\end{align}
Take a pure state in the Hilbert
space $\ket{\Psi}$. 
We denote by $\MM$ the superoperator corresponding to filtering
out the state $\ket\Psi$, i.e., 
\begin{align}
\MM[\rho] = (\II-\ket{\Psi}\bra{\Psi})
\rho
(\II-\ket{\Psi}\bra{\Psi}).  
\end{align}

We introduce a shorthand for (unnormalized) pure states obtainable
from $\ket\Psi$ via the operators $A_j$. For each sequence
$\jj=(j_1,\ldots,j_t)$ of integers $j_n \in [1,D]$, we define
\begin{align}
\ket{\underline{j}} &= 
A_{j_t} \cdots A_{j_2}
A_{j_1} \ket{\Psi};
\label{eq:pure_states_t}
\\
p_{\underline{j}}&= 
\braket{\jj}{\jj}.
\end{align}
The $t$th iterate of $\ket{\Psi}\bra\Psi$ under $\SSS$ can be written with
these states as
\begin{align} 
\SSS^t[\ket{\Psi}\bra{\Psi}] &= \sum_{j_1=1}^D \sum_{j_2=1}^D \ldots
\sum_{j_t=1}^D \ket{\underline{j}}
\bra{\underline{j}}.
\end{align} 
This is a probabilistic mixture of the pure states
$\ket{\jj}$ with weights $p_{\jj}$. 

Similarly, we use $\ket{\jj}_\text{cond}$ to denote 
(unnormalized) pure states obtainable
from $\ket\Psi$ via the operators $(\II-\ket{\Psi}\bra\Psi)A_j$, 
\begin{align}
\ket{\underline{j}}_\text{cond} &= (\II-\ket{\Psi}\bra{\Psi})
A_{j_t}\cdots (\II-\ket{\Psi}\bra{\Psi}) A_{j_1} \ket{\Psi}
\nonumber\\ \quad &= \ket{\jj}+\sum_{n=2}^{t+1}
c_{(j_n,\ldots,j_t)}(\jj) \ket{j_n,\ldots,j_t},
\label{eq:cond_states_t}
\end{align}
where the sequences $(j_n,\ldots,j_t)$ are obtained from the sequence
$\jj=(j_1,\ldots,j_t)$ by omitting the first $t-1$ elements (including
the case $n=t+1$, where we obtain the empty sequence, for which
according to Eq.~\eqref{eq:def_rel}, $\ket{\emptyset}=\ket{\Psi}$). The
coefficients $c_{(j_n,\ldots,j_t)}(\jj) \in \mathbb{C}$ are complex
numbers. 
The $t$th iterate of $\ket\Psi\bra\Psi$ under $\MM \SSS$ can be written
using these states as 
\begin{align} 
(\MM \SSS)^t[\ket{\Psi}\bra{\Psi}] &= \sum_{j_1=1}^D
  \sum_{j_2=1}^D \ldots \sum_{j_t=1}^D \ket{\underline{j}}_\text{cond}
  \bra{\underline{j}}_\text{cond}.
\end{align} 

The \emph{relevant Hilbert space} $\Hrel(\SSS,\ket\Psi)$ is the
space spanned by the vectors $\ket{\jj}$, for all admissible sequences
$\jj$ of any legth $t\in\mathbb{N}$. 
The projector to this subspace of $\mathcal{H}$ is the limit
\begin{align}
\Irel(\SSS,\ket{\Psi})
&= \lim_{n\to\infty} \PPP(\sum_{t=0}^n \SSS^n[\ket{\Psi}\bra{\Psi}]),
\end{align}
where $\PPP(\sigma)$ denotes the projector to the nonzero subspace
(support) of a Hermitian operator $\sigma$.  

It is clear by the construction of the set $\{\ket{\jj}\}$ that the
relevant Hilbert space is the smallest invariant subspace of $\SSS$
that contains $\ket{\Psi}$. It is an invariant subspace, since if
$\sigma$ is a density operator in $\Hrel(\SSS,\ket\Psi)$, then it can
be decomposed as $\sigma=\sum_{\jj} r_{\jj} \ket{\jj}\bra{\jj}$, and
then $\SSS^n[\sigma]$ is also in $\Hrel(\SSS,\ket\Psi)$, for any $n\in
\mathbb{N}$. On the other hand, it contains $\ket\Psi$, and is the
smallest such subspace, since it does not contain any state $\ket\Phi$
that is not reachable from $\ket\Psi$ by iterations of $\SSS$. Indeed,
for such states $\ket\Phi$, we would have $\bra{\Phi}\SSS^t[\ket\Psi
  \bra\Psi]\ket{\Phi}=0$ for all $t\in \mathbb{N}$, and thus, they
would be outside of $\Hrel(\SSS,\ket\Psi)$.

We now show that the relevant Hilbert space is also spanned by the
vectors $\ket{\jj}_\text{cond}$, i.e., that 
\begin{align}
\Irel(\SSS,\ket{\Psi})
&= 
\Irel(\MM \SSS,\ket{\Psi}). 
\label{eq:m_s}
\end{align}
First, from the second line of Eq.~\eqref{eq:cond_states_t}, every
vector ${\ket\jj}_\text{cond}$ is expressed as a linear combination of
vectors $\ket{\jj}$, and so, 
$\Irel(\MM \SSS,\ket{\Psi}) \le
\Irel(\SSS,\ket{\Psi})$. 
On the other hand, every vector $\ket{\jj}$
can be expressed as linear combination of $\ket\jj_\text{cond}$ and of
vectors $\ket{\jj'}_\text{cond}$, where the $\jj'$ are sequences
shorter than $\jj$. This can be shown using mathematical induction,
started from sequences of length $t=1$, for which
\begin{align}
\ket{(j_j)} &= \ket{(j_1)}_\text{cond} + \ket{\Psi}\braket{\Psi}{(j_1)},
\label{eq:cond_states_1}
\end{align}
and using Eq.~\eqref{eq:cond_states_t} for the inductive step.  Thus,
$\Irel(\SSS,\ket{\Psi}) \le \Irel(\MM \SSS,\ket{\Psi}) $, and
this, together with $\Irel(\MM \SSS,\ket{\Psi}) \le
\Irel(\SSS,\ket{\Psi})$ shown above, proves Eq.~\eqref{eq:m_s}.

The results of this Appendix translate to DTQWs considered in the
paper, and prove Eq.~\eqref{eq:def_rel}.

\section{Numerical methods for the expected return time}
\label{app:numerics}

To study the expected return time numerically in more detail, we use
the spectral decomposition of the conditional timestep operator
$\MM\SSS$. This method is applicable only if the matrix of $\MM\SSS$
is diagonalizable, which is the generic case.

We first find all eigenstates $\chi_n$ of $\MM\SSS$, for
$n=1,\ldots,M^2$, with  eigenvalues
$\alpha_n\in\mathbb{C}$,
\begin{align}
\forall n: \quad \MM\SSS[\chi_n]&=\alpha_n \chi_n.
\end{align}
The next step is to provide a decomposition of the initial state
$\ket{\Psi}\bra{\Psi}$ in terms of the eigenstates $\chi_n$, 
\begin{align}
\ket{\Psi}\bra{\Psi} &= \sum_n c_n \chi_n.
\end{align}
The coefficients $c_n\in \mathbb{C}$ in the decomposition can be found
using the right eigenvectors of the matrix of $\MM\SSS$.  Using the
above, we can write the expected return time as a geometric series,
and obtain
\begin{align}
\TPsi &= \text{Tr } \trr = 
\sum_n \frac{c_n}{1-\alpha_n} \text{Tr }\chi_n. 
\label{eq:numeric_exact}
\end{align}
To study convergence of the expected return time, we define the
expected return time up until a finite number $L$ of timesteps, as
\begin{align}
\TPsi^{(L)} &= \sum_{t=1}^L t p_t + (L+1) (1-\sum_{t=1}^L p_t) = \sum_{t=0}^L q_t,
\end{align}
efficiently calculated using the spectral decomposition as 
\begin{align}
\TPsi^{(L)} &= \text{Tr } \sum_{t=0}^L (\MM\SSS)^t[\ket{0}\bra{0}] = 
\sum_n c_n \frac{\alpha_n^{L+1}-1}{\alpha_n-1} \text{Tr }\chi_n.
\end{align}

\bibliography{returntime}

\end{document}